\documentclass[journal=jacsat,manuscript=article]{achemso}

\usepackage{color}
\usepackage{amsmath}
\usepackage{pifont}
\usepackage{amssymb}  
\usepackage{bbold}
\usepackage{float}
\usepackage{tikz}
\usepackage{makecell}
\usepackage{subfigure}
\usepackage{pifont}   
\usepackage{graphicx} 
\usepackage{dcolumn}  
\usepackage{bm}       
\usepackage{multirow} 
\usepackage{hyperref} 
\usepackage{placeins}

\author{Martin Rodriguez-Vega}
\affiliation{Theoretical Division, Los Alamos National Laboratory, Los Alamos, New Mexico 87545, USA}

\author{Ze-Xun Lin}
\affiliation{Department of Physics, The University of Texas at Austin, Austin, TX 78712, USA}
\alsoaffiliation{Department of Physics, Northeastern University, Boston, MA 02115, USA}

\author{A. Leonardo}
\affiliation{Donostia International Physics Center, Paseo Manuel de Lardizabal 4, 20018 San Sebastian, Spain}
\alsoaffiliation{Department of Physics, University of the Basque Country UPV/
EHU, Leioa, Spain.}

\author{A. Ernst}
\affiliation{Institut f\"ur Theoretische Physik, Johannes Kepler Universit\"at, A 4040 Linz, Austria}
\alsoaffiliation{Max-Planck-Institut f\"ur Mikrostrukturphysik, Weinberg 2, D-06120 Halle, Germany}

\author{M. G. Vergniory}
\affiliation{Donostia International Physics Center, Paseo Manuel de Lardizabal 4, 20018 San Sebastian, Spain}
\alsoaffiliation{Max Planck Institute for Chemical Physics of Solids, Dresden, D-01187, Germany}

\author{Gregory A. Fiete}
\affiliation{Department of Physics, Northeastern University, Boston, MA 02115, USA}
\alsoaffiliation{Department of Physics, Massachusetts Institute of Technology, Cambridge, MA 02139, USA}

\title{Light-driven Topological and Magnetic Phase Transitions in Thin-layer Antiferromagnets}

\begin{document}

\date{\today}

\begin{abstract}
  We theoretically study the effect of low-frequency light pulses in
  resonance with phonons in the topological and magnetically ordered
  two septuple-layer (2-SL) MnBi$_2$Te$_4$ (MBT) and MnSb$_2$Te$_4$
  (MST). These materials share symmetry properties and an
  antiferromagnetic ground state in pristine form but present
  different magnetic exchange interactions. In both materials, shear
  and breathing Raman phonons can be excited via non-linear
  interactions with photo-excited infrared phonons using intense laser
  pulses attainable in current experimental setups. The light-induced
  transient lattice distortions lead to a change in the sign of the
  effective interlayer exchange interaction and magnetic order
  accompanied by a topological band transition. Furthermore, we show
  that moderate anti-site disorder, typically present in MBT and MST
  samples, can facilitate such an effect. Therefore, our work
  establishes 2-SL MBT and MST as candidate platforms to achieve
  non-equilibrium magneto-topological phase transitions.
\end{abstract}
\maketitle

\section{Graphical TOC Entry}

\begin{figure}
  \centering
  \includegraphics[width=2in]{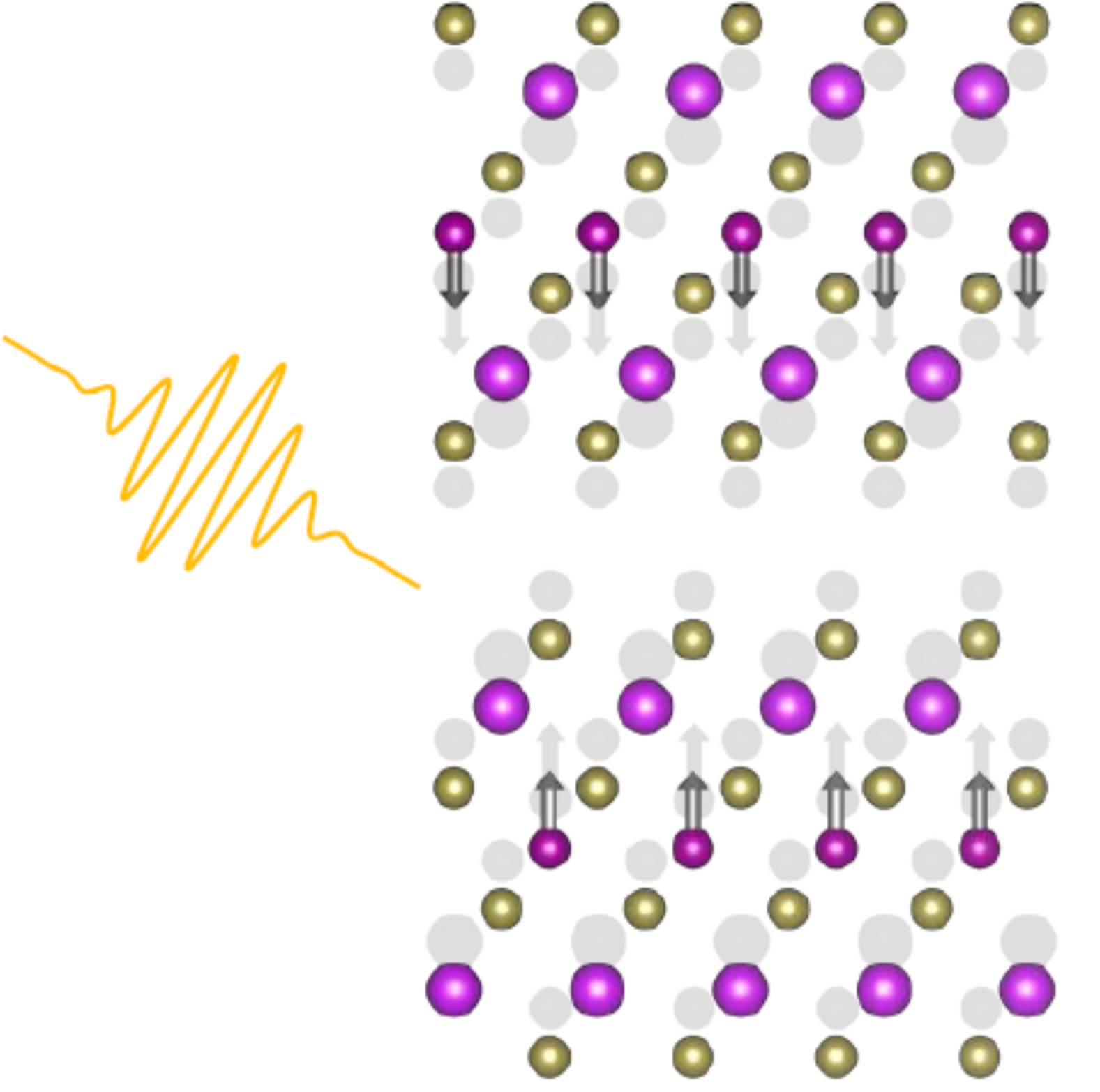}
\end{figure}
\clearpage

Antiferromagnetic topological insulators (ATIs) can host exotic phases
of matter such as the quantum anomalous Hall (QAH) effect and axion
insulators~\cite{PhysRevB.81.245209}. The search for these topological
phases motivated the addition of magnetic dopants in topological
insulators, which led to the observation of a QAH effect and candidates for
axion insulators at very low
temperatures~\cite{chang_experimental_2013,Mogi2017,Gooth2019}.
However, intrinsic ATIs promise to manifest these phases at higher
temperatures, desirable for applications. Indeed, the recent
predictions, synthesis and exfoliation of the van der Waals materials
MnBi$_2$Te$_4$, MnBi$_{2n}$Te$_{3n+1}$, and
MnSb$_2$Te$_4$~\cite{otrokov_prediction_2019,gong_experimental_2019,hu_van_2020,hu_realization_2020,jahangirli_electronic_2019,aliev_novel_2019,ge2021direct,Xu2020high-throughput}
allowed the detection of QAH states in odd septuple layers (SLs),
axion states in even SLs ~\cite{liu_robust_2020,
  deng_quantum_2020,ovchinnikov2020intertwined,10.1093/nsr/nwaa089,lupke2021local},
and the observation of an electric-field-induced layer Hall effect in
six SL samples~\cite{Gao2021}.

The intertwined nature of the magnetic and topological order in ATIs
offers the possibility to explore topological transitions induced by
changes in the magnetic order and vice-versa. For example, recent
experiments suggest that increasing the distance between the magnetic
planes in the MnBi$_{2n}$Te$_{3n+1}$ family leads to ferromagnetic
order~\cite{hu_realization_2020}. On the other hand, reducing the
distance in MnBi$_2$Te$_4$ single crystals via hydrostatic pressure
leads to the suppression of the AFM
order~\cite{Chen_2019_supp,CuiyingPei66401}. In contrast, in CrI$_3$ -
a low-dimensional magnetic system with trivial topology - hydrostatic
pressure induces an antiferromagnet (AFM) to ferromagnet (FM)
transition~\cite{Li2019CrI3pressure}. However, a suitable mechanism
to modify the magnetic order in ATIs without applied external magnetic
fields or superlattices remains elusive.

To this end, non-equilibrium approaches provide a possible pathway to
achieve magneto-topological transitions in
ATIs~\cite{Mankowsky_2016,Oka_2019,rudner2020_review,RODRIGUEZVEGA2021168434}.
Most notably, non-linear phononics~\cite{forst2011,subedi2014,Mankowsky_2016} -- a
transient and controlled lattice distortion induced by photo-excited
phonons-- has been successfully used to transiently enhanced
superconductivity~\cite{Fausti2011,Mankowsky2014,mitrano2016}, manipulate and induce 
ferroelectric states~\cite{Mankowsky2017,Nova1075}, and induce dynamical ferrimagnetic transitions \cite{Disa2020}. 
More recently, A. Stupakiewicz {\it et al.} induced switching of magnetization in yttrium
iron garnet (YIG) thin films by pumping of phonon
modes~\cite{Stupakiewicz2021}. More generally, light has been shown to induce metastable charge-density-wave states~\cite{Vaskivskyie1500168} and incite
transitions into hidden phases~\cite{Stojchevska2014}. This experimental evidence motivates the
use of non-equilibrium approaches to manipulate magneto-topological
order in ATIs. In this work, we show theoretically that an AFM to FM magnetic
transition accompanied by a topological transition can be induced in
2-SL MXT (X=Bi, Sb) samples with intense, experimentally accessible THz laser pulses in
resonance with the phonons.
Interestingly, the moderate anti-site disorder typically present in
these materials reduces the laser intensity threshold to induce the
transition.

\begin{figure}
  \centering
  \includegraphics[width=8.5cm]{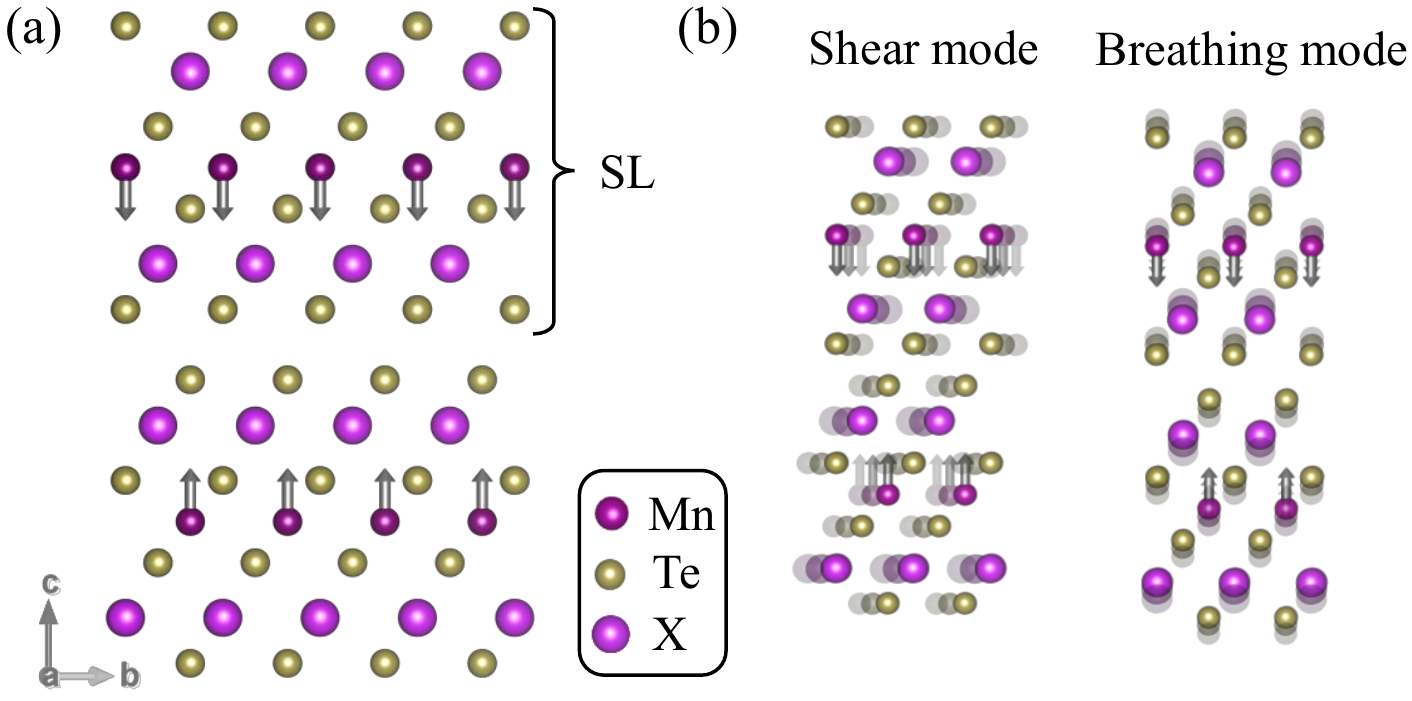}
  \caption{(Color online) (a) 2-SL MXT lattice structure and magnetic
    order (moments shown in gray arrows). X = Bi, Sb atoms are represented in pink, Te
    atoms in yellow, and Mn atoms in purple. (b) Low-frequency shear
    and breathing modes characteristic of few-layer materials. The
    breathing mode preserves all the crystal symmetries.}
\label{fig:fig1}
\end{figure}

In MXT materials, the constituent SLs (see Fig. \ref{fig:fig1}(a)) are
held together via van der Waals forces, which allows exfoliation in
thin samples~\cite{Deng_2020,Liu2020mbt}. We will focus on systems
with two SLs, since they correspond to the minimal system which can
accommodate interlayer AFM order. Within each layer, the magnetic
moments are aligned ferromagnetically, but opposite layers posses
opposite magnetic moment directions. For 2-SL MBT, the critical
temperature is approximately $20$~K~\cite{PhysRevX.11.011003}. For
bulk MST, a critical temperature of $19$~K has been
reported~\cite{YanMST2019}. However, depending on synthesis conditions, bulk MST can posses a
ferromagnetic ground state~\cite{wimmer2021mnrich,Wenbo2021}.

2-SL MBT and 2-SL MST present space group $P\bar3m1$ (No. 164) with
point group $D_{3d}$ in their paramagnetic phase. The unit cell
contains $N=14$ atoms with Te atoms located in Wyckoff positions
$2d$ $(1/3,2/3,z)$ and $2c$ $(0,0,z)$, Mn atoms at $2d$, and X=Bi, Sb
atoms in $2c$ and $2d$ positions. The lattice vibration representation is given by $\Gamma_{vib} = 7
A_{1g} \oplus 7 A_{2u} \oplus 7 E_g \oplus 7 E_u$, which corresponds
to 7 non-degenerate ($A_{1g}$) and 7 double-degenerate ($E_g $) Raman
modes, with equal number of their infrared counterparts, including the
three acoustic modes ($E_u \oplus A_{2u}$). The character table for
$D_{3d}$ is shown in the Supporting Information. 

Employing group theory and projection
operators, we derive the set of real-space displacements which bring the
dynamical matrix into block-diagonal form, according to their
irreducible representations (see Methods for details). We find that the shear
mode where one SL shifts in the $[100]$ direction and the opposite SL
in the $[\bar 100]$ direction belongs to the $E_g$ irrep. Its partner
corresponds to an orthogonal in-plane displacement. The breathing mode
consists of the SLs displacing away and towards each other in the
direction normal to the plane ($[001]$ and $[00\bar1]$ respectively),
and belongs to the $A_{1g}$ representation. Fig. \ref{fig:fig1} (b)
shows representations of these modes.  For a detailed group theory
study of few-SLs MBT, see Ref. ~\cite{Rodriguez-Vega2020_mbt}.

\begin{figure}
  \centering
  \includegraphics[width=8.5cm]{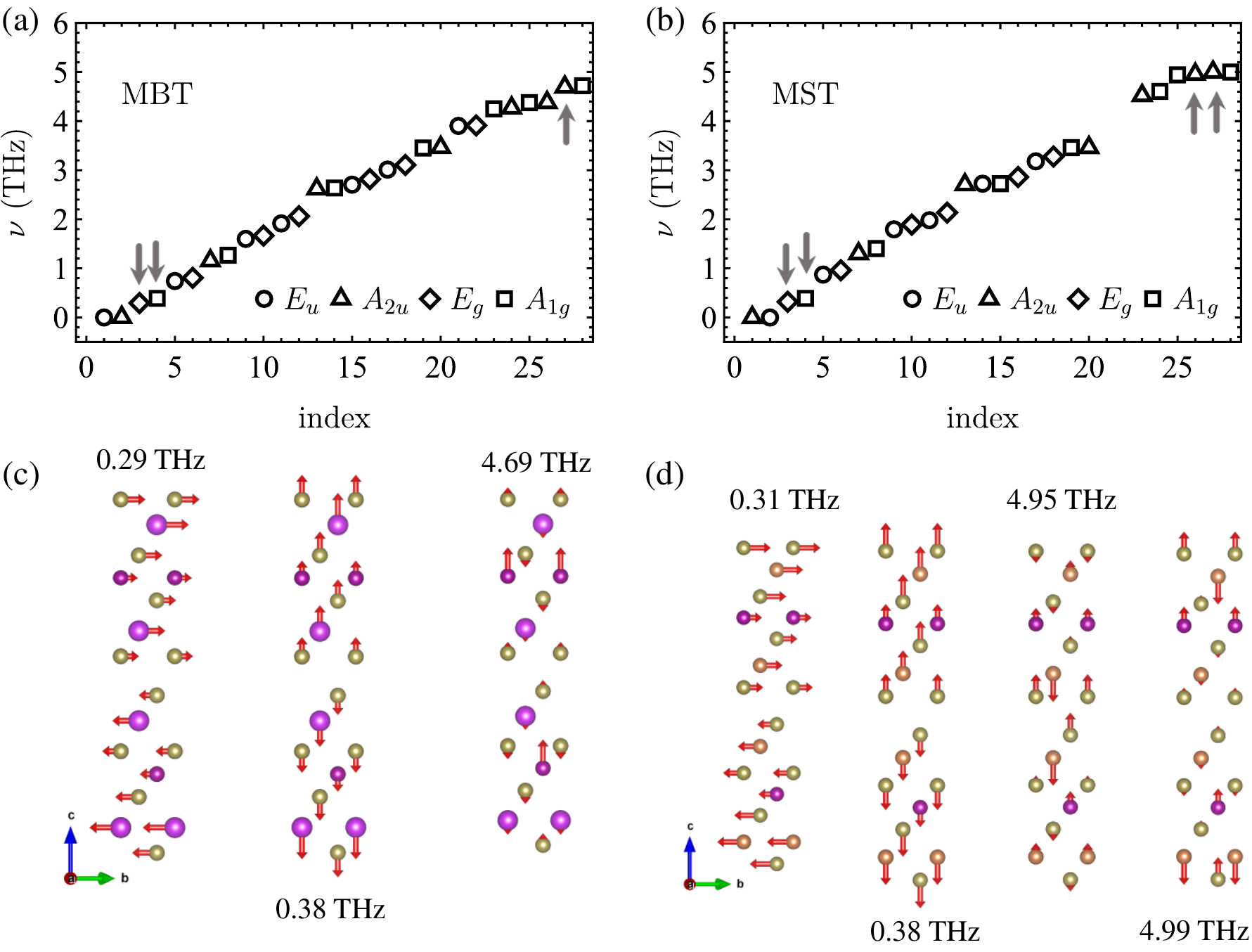}
  \caption{(Color online.) Phonon frequencies for (a)
    2-SL MBT and (b) 2-SL MST obtained with
    first-principles calculations. The gray arrows
    indicate the phonons illustrated below. In panels
    (c) and (d) we show the real-space lattice
    displacements with their corresponding
    frequencies. Red arrows indicates the
    displacements.}
\label{fig:freqs}
\end{figure}

Now that we have established that the shear and breathing modes are
allowed by symmetry, and determined their irreps, we calculate the
phonon frequencies at the $\Gamma$ point.  We considered paramagnetic, FM
and AFM configurations without spin-orbit coupling and find only
negligible differences among the corresponding phonon frequencies. The results for both 2-SL MBT and MST are summarized in Fig.
\ref{fig:freqs}. Panels (a) and (b) show the $\Gamma$-point phonon
frequencies with their corresponding irreducible representation
indicated by the shape of the marker. In both materials, the shear and
breathing modes present the smallest frequency among the optical modes
(indicated by downward gray arrows), and their frequency is smaller by
a factor of two compared with the next optical phonon.

Having characterized the properties of the phonons in the harmonic regime, we next consider the symmetry aspects of their non-linear interactions and their laser excitation. A laser pulse incident onto a sample can couple directly with IR
modes, depending on the laser frequency and electric field direction.
In turn, such an IR mode can couple non-linearly with some Raman
modes, provided their irreps satisfy
$\left[\Gamma_{\mathrm{IR}} \otimes \Gamma_{\mathrm{IR}}\right]
\otimes \Gamma_{R} \supset A_{1 g}$~\cite{subedi2014}. This mechanism
is referred to as non-linear
phononics~\cite{forst2011,FORST201324,subedi2014}, and has allowed
experimental~\cite{Fausti2011,Mankowsky2014,mitrano2016,Mankowsky_2016,FORST201324,Nova1075,Nova2017}
and theoretical manipulations of correlated states of
matter~\cite{Sentef2016,kalsha2018,juraschek2017,PhysRevB.102.081117}. For the 2-SL MXT's point group, driving a
$A_{2u}$ mode can rectify totally-symmetric modes, such as the
breading modes, since $A_{2u} \otimes A_{2u} = A_{1g}$. Thus, the
shear modes ($E_g$ irrep) are not affected. On the other hand, driving
an $E_u$ mode allows coupling with the low-frequency shear modes in
conjunction with the breading mode, since
$E_u \otimes E_u = A_{1 g} \oplus A_{2 g} \oplus E_{g}$.

Once an IR mode has been driven with a strong-enough laser pulse,
coupling to all Raman modes with compatible irreps is allowed by
symmetry. However, in our case, since the solution of the dynamical
equations scale with the inverse square of the Raman frequency
($\sim \Omega_R^{-2}$), we can simplify the calculation and restrict
the non-linear interactions to only the low-frequency shear and
breathing modes~\cite{subedi2014,juraschek2017}. We now consider a laser pulse optimized to couple with the
highest-frequency IR modes, with irrep $A_{2u}$. This
mode presents the strongest coupling with the laser as shown by the
largest Born effective charge $\boldsymbol{Z^*}$~\cite{gonze1997,
  baroni2001} (see the Supporting Information). In this case, the non-linear potential
for 2-SL MBT takes the form
\begin{align}
&V[Q_{IR},Q_{\text{R}(3)},t] =  \frac{1}{2}\Omega^2_{\text{IR}} Q_{\text{IR}}^2+ \frac{1}{2}\Omega^2_{\text{R}(3)} Q^{2}_{\text{R}(3)}+\gamma_3 Q_{\text{IR}}^2 Q_{\text{R}(3)} + \frac{1}{3}\beta_3 Q^{3}_{\text{R}(3)} + \boldsymbol{Z^*} \cdot \boldsymbol{E_0} \sin(\Omega t) F(t) Q_{\text{IR}},
\label{eq:non-linear-pot_mbt_a2u_a}
\end{align}
where $\gamma_3$ and $\beta_3$ are non-linear coefficients determined
from DFT calculations (for the procedure and numerical values, see
the Supporting Information), $\boldsymbol{E_0}$ is the electric field amplitude with
Gaussian profile $F(t)=\exp\{ -t^2/(2 \tau^2)\}$, and $\Omega$ is the
laser frequency, which we choose in resonance with the IR mode $\Omega
=\Omega_{IR} = 4.69$~THz. Notice that the driven $A_{2u}$ non-linear potential is
much simpler than the one for driven $E_u$ modes. This is because the $A_{2u}$ phonons
do not couple to $E_g$ modes up to cubic order interactions. 

For 2-SL MST, there are two IR modes with $A_{2u}$ irreps, similar
Born effective charge, and frequency. Therefore, we need to consider
the simultaneous excitation of the two $A_{2u}$ IR modes which leads
to the potential
\begin{align}\nonumber
&V[\{ Q_{IR(i)} \},Q_{\text{R}(3)},t] =  \sum_{i=1}^2 \frac{1}{2}\Omega^2_{\text{IR}(i)} Q_{\text{IR}(i)}^2+ \frac{1}{2}\Omega^2_{\text{R}(3)} Q^{2}_{\text{R}(3)}+\\
&\gamma_{1,3} Q_{\text{IR}(1)}^2 Q_{\text{R}(3)} +
\gamma_{2,3} Q_{\text{IR}(2)}^2 Q_{\text{R}(3)} +
\gamma_{1,2,3} Q_{\text{IR}(1)} Q_{\text{IR}(2)} Q_{\text{R}(3)} \nonumber \\
&+\frac{1}{3}\beta_3 Q^{3}_{\text{R}(3)} + \left( \boldsymbol{Z_1^*}Q_{\text{IR}(1)} + \boldsymbol{Z_2^*}Q_{\text{IR}(2)} \right) \cdot \boldsymbol{E_0} \sin(\Omega t) F(t).
\label{eq:non-linear-pot_mbt_a2u}
\end{align}

For 2-SL MST, we consider the laser frequency $\Omega =(\Omega_{IR(1)}
+\Omega_{IR(2)})/2$~THz. The phonon dynamics are determined by the
equations of motion $\partial^2_t Q_{\text{R}}  = -\partial_{Q_{\text{R}}} V[Q_{\text{IR}(i)},Q_{\text{R}}]$, $\partial^2_t Q_{\text{IR}(i)} = -\partial_{Q_{\text{IR}(i)}} V[Q_{\text{IR}(i)},Q_{\text{R}}]$, where $i$ runs over the driven IR modes. We solve the differential
equations numerically. In this work, we do not consider the phonon
lifetime. Recent Raman measurements have shown that the lifetime of the breathing mode is approximately $13.3$ ps \cite{Choe2021}, which is sufficiently long for the electronic degrees of freedom to respond.

\begin{figure}
  \centering
  \includegraphics[width=8.5cm]{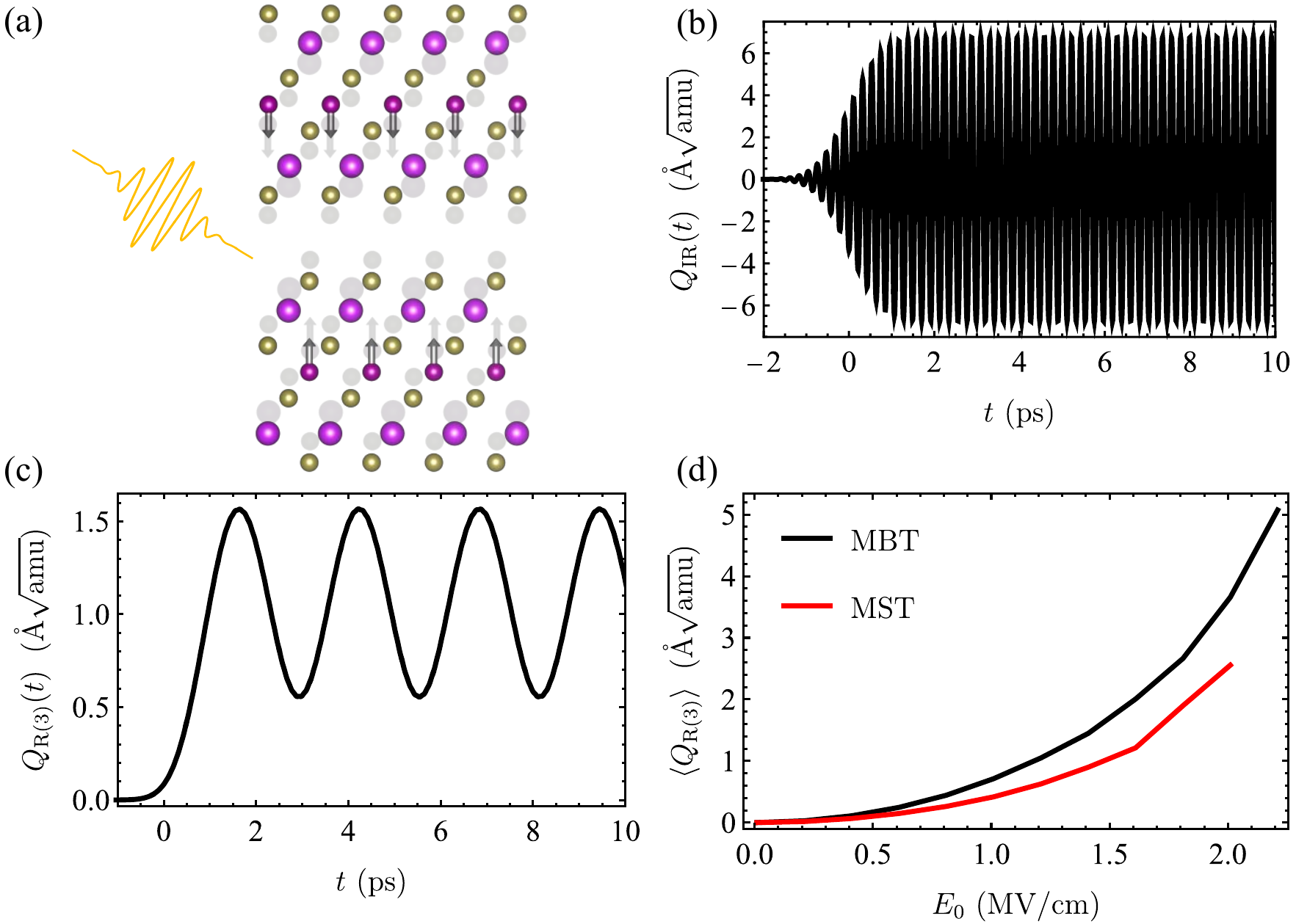}
  \caption{(Color online.) (a) Sketch of a light-induced lattice
    distortion. (b) Time dependence of the infrared phonon mode
    directly excited by the incident laser pulse in 2-SL MBT. (c)
    Non-linearly-excited breathing mode, which oscillates about a new
    shifted position. The laser parameters used in (b) and (c) are
    $\tau=0.6$~ps and $E_0 = 0.6$~MV/cm. (d) Average displacement of
    the non-linearly photo-excited breathing mode $Q_{\text{R}(3)}$
    for MBT (black) and MST (red) for $\tau = 0.3$~ps and laser
    frequency $\Omega = \Omega_{\text{IR}(1)}$ for 2-SL MBT and
    $\Omega = (\Omega_{\text{IR}(1)}+\Omega_{\text{IR}(2)})/2$ for
    2-SL MST.}
\label{fig:fig3}
\end{figure}

%
The phonon dynamics for a general laser intensity and pulse
duration can be obtained by solving the equations of motion
numerically. In Fig. \ref{fig:fig3}(a), we show a sketch of a
laser-irradiated 2-SL MBT sample. The incoming light with frequency
$\Omega = \Omega_{\text{IR}}=4.69$~THz couples directly to the
corresponding resonant IR mode. As we show in Fig.  \ref{fig:fig3}(b),
this mode oscillates around its equilibrium position. Anharmonic
coupling induces dynamics in the Raman breathing mode, even though it
does not couple directly to the laser. The non-linear nature of the
interaction ($\gamma_3 Q_{\text{R}} Q^2_{\text{IR}}$) leads to oscillations
about a position shifted with respect to the equilibrium position. 
 Fig.  \ref{fig:fig3} (c) we shows such oscillations for a laser with
peak electric field $E_0 = 0.6$ and pulse duration $\tau =
0.6$~ps. Similar responses are obtained in 2-SL MST, where the main
difference is the presence of two $A_{2u}$ IR modes, instead of
one. Notice that with light, we can only obtain
$\langle Q_{\text{R}(3)} \rangle \ge 0$, which corresponds to an
effective increase in the Mn-Mn layer separation. This a consequence of 
the sign of the non-linear coefficients ($\gamma_3$ for MBT, $\gamma_{1,3},\gamma_{2,3}$ for MST)--intrinsic for the materials.

The complementary
process of bringing the Mn planes closer to each other could be
achieved by applying uniaxial pressure. Theoretically,
Ref.~\cite{10.1088/1367-2630/ac1974} predicts that bulk MBT undergoes
a topological quantum phase transition under 2.12\% compressive strain.

In Fig \ref{fig:fig3}(d), we plot the time average of the shear modes
as a function of $E_0$ for $\tau=3$ ps. Experimentally, fields of up
to $100$~MVcm$^{-1}$ have been reported in the range $15-50$
THz~\cite{sell2008,kampfrath2013} but limitations are imposed by the
amplitude of the corresponding lattice distortion. For 2-SL MBT (2-SL
MST) , $\langle Q_{\text{R}(3)} \rangle = 5 $ \AA/$\sqrt{\text{amu}}$
correspond to $1.68 \%$ ($1.88 \%$) increase in the Mn-Mn plane
interlayer distance. For 2SL-MST, the dynamical equations become
unstable for $E_0 \gtrapprox 2$~MV/cm. However, the range of stability is large enough to obtain a
magnetic transition.

Inelastic neutron scattering measurements
\cite{Li_2020_mbtspin} suggest that the magnetic order in bulk MBT is
described by the local-moment Hamiltonian ($S=5/2$)
$ 
  \mathcal{H} = \mathcal{H}_{intra} + \mathcal{H}_{inter},  
$ 
where the intralayer Hamiltonian can be written as
$ 
   \mathcal{H}_{intra} = - \sum_{ij} J_{ij} \boldsymbol{S}_i \cdot \boldsymbol{S}_j-D \sum_i \left( S^z_i \right)^2, 
$ 
with exchange interaction $J_{ij} $ (up to fourth-neighbor
interactions are needed to fit the data correctly with $SJ_1=0.3$~meV,
$SJ_2=-0.083$~meV, and $SJ_4=0.023$~meV), and $SD=0.12$~meV is a
single-ion anisotropy. Thus, the effective intralayer coupling is
positive and leads to the ferromagnetic order in each Mn layer. 
The interlayer Hamiltonian is given by
$ 
\mathcal{H}_{inter} = - J_c \sum_{ \langle ij \rangle } \boldsymbol{S}_i \cdot \boldsymbol{S}_j,  
$ 
where experiments suggest a nearest-neighbor AFM interlayer
interaction $SJ_c=-0.055$~meV~\cite{Li_2020_mbtspin}. We obtain the
spin Hamiltonian from first-principle calculations, employing a
Green's function approach and the magnetic force
theorem~\cite{Liechtenstein1987,hoffmann2020}. The calculations were
performed using a GGA+U approximation, which describes adequately
localized Mn $3d$ states with $U_{eff}=U-J$=5.3
eV~\cite{otrokov_prediction_2019,wimmer2021mnrich}.  For the
interlayer interactions, the Hamiltonian takes the more general form
$ \mathcal{H}_{inter} = - J_c \sum_{ ij } \boldsymbol{S}_i \cdot
\boldsymbol{S}_j $, where longer-range interactions are relevant. In
pristine MXT compounds the interlayer coupling governs the
antiferromagnetic order in the ground state, which is mainly mediated
by a long-range double exchange interaction via Te
ions~\cite{otrokov_prediction_2019,wimmer2021mnrich}. However, natural lattice
defects such as antisite Mn-Bi or Mn-Sb disorder or Mn excess in
Bi (Sb) layers can lead to ferromagnetic order in these
systems~\cite{wimmer2021mnrich, Lai_2021}. 


We now study the effect of laser-induced transient
lattice distortions on the magnetic order.  Under a time-dependent
lattice deformation, small compared with the equilibrium inter-atomic
distances, the spin exchange interaction can be approximated
as~\cite{granado1999},
\begin{equation}
J[ \boldsymbol{u}(t) ] = J^0 +  \delta J \boldsymbol{\hat \delta} \cdot   \boldsymbol{u}(t) + \mathcal O(\boldsymbol{u}(t)^2),
\end{equation}
where $\boldsymbol{u}(t)$ is the real-space lattice displacement,
$J^0$ is the equilibrium interaction, and $\delta J$ is the coupling
constant between the phonon and the spins. The connection with the
phonon amplitude $Q$ is given by $ \boldsymbol{u}_{\kappa} =
{Q}/{\sqrt{m_\kappa}} \boldsymbol{e}_{\kappa}, $ where $m_\kappa$ is
the mass of atom $\kappa$, and $\boldsymbol{e}_{\kappa}$ are the
normalized dynamical matrix eigenvectors.

Next, we define the effective spin interaction employing Floquet
theory. The exchange interactions set the relevant energy scale, with
$\lesssim 1$~meV. Since the infrared phonon frequency
($\Omega_{\text{IR}} \approx 4.95, 4.69$~THz) used is larger than the
exchange energy, we can define an effective time-averaged exchange
interaction
$ 
J^{\text{eff}} = J^0 +  \delta J \boldsymbol{\hat \delta} \cdot \langle \boldsymbol{u}_{\text{R}} \rangle,
$ 
where $\langle \cdots \rangle$ indicates the time average. Thus, when
the phonons oscillate about their equilibrium positions (harmonic
phonons), such that $\langle \boldsymbol{u} \rangle = 0$, the exchange
interactions are not modified in the picture discussed here. The
non-zero average shift, however, can renormalize the interactions
leading to different magnetic configurations compared with the
equilibrium counterparts.

\begin{figure}
  \centering
  \includegraphics[width=7.0cm]{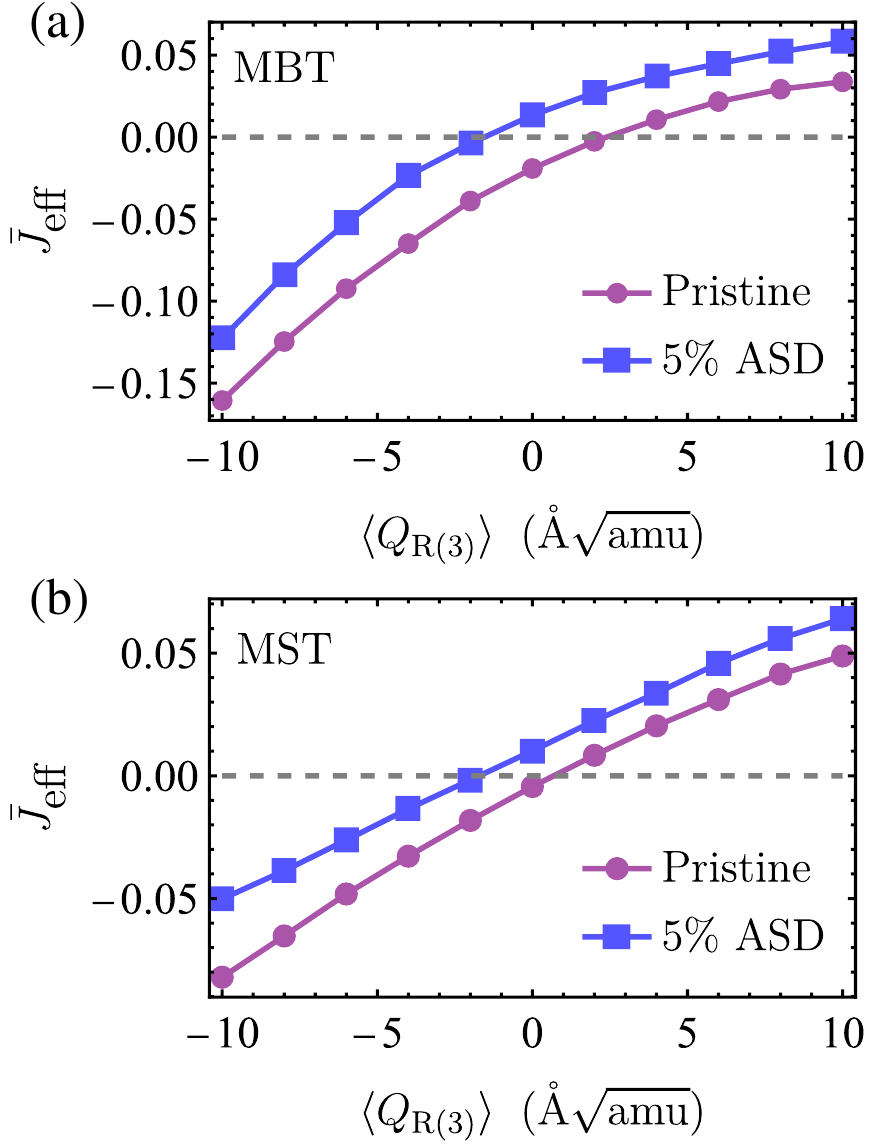}
  \caption{(Color online.) Effective averaged interlayer exchange
    interaction as a function of the average breathing mode $\langle
    Q_{\text{R}(3)} \rangle$ for (a) 2SL-MBT and (b) 2SL-MST. The
    purple circles correspond to pristine samples, while squares
    correspond to $5 \%$ anti-site disorder (ASD).}
\label{fig:fig4}
\end{figure}

We compute the light-induced effective exchange interactions as a
function of the phonon amplitude $Q_{\text{R}(3)}$. Our results are
summarized in Fig. \ref{fig:fig4}. We plot the average interlayer
exchange interaction
$\bar{J}_{\text{eff}} = 1/\mathcal{N}\sum_{ij} J_{ij}$ as a function
of $Q_{\text{R}(3)}$. We used a supercell, which consists of seven SLs 
of MBT (MST) and three SLs of vacuum simulated by empty spheres. $\bar{J}_{\text{eff}}$ 
represents an average exchange interaction, and $\mathcal{N}$ is the number of interacting magnetic 
moments taken for the average. For pristine 2-SL MBT (2-SL MST), we find a sign
change in the interlayer exchange interaction at
$Q_{\text{R}(3)} \approx 2.4$ \AA $\sqrt{\text{amu}}$
($Q_{\text{R}(3)} \approx 0.7$ \AA $\sqrt{\text{amu}}$). These phonon
amplitudes can be obtained with a laser pulse with $E_0 \approx 1.7$
MV/cm and $\tau=0.3$~ps ($E_0 \approx 1.5$ MV/cm and $\tau=0.3$~ps),
as we show in Fig. \ref{fig:fig3}. Generally, increasing the
vertical distance between the Mn magnetic moments weakens the
antiferromagnetic coupling and favours ferromagnetic order in
these systems. The time scale for the spin reorientation following
the sign change in $\bar{J}_{\text{eff}}$ depends on parameters such as
the Gilbert damping factor \cite{1353448}, the exact spin anisotropy for 2-SL
MBT and MST, and the laser-induced $\bar{J}_{\text{eff}}$, but are within the limits of the effect we predict to occur.


Since MXT samples are prone to anti-site disorder
\cite{yan_crystal_2019,PhysRevX.11.021033,Yuan2020,C9CP05634C,Lai_2021,wimmer2021mnrich},
with disorder percentages dependent on the sample fabrication process,
we also discuss the role of disorder in the light-induced magnetic
transition. Depending on concentration percentage, Mn-Sb anti-site
disorder can tune the interlayer magnetic interaction into
ferromagnetic states \cite{PhysRevX.11.021033}. Here, we study
theoretically the role of the anti-site disorder in the light-induced
magnetic transition discussed in the previous section.

First, we will assume that the anti-site disorder has a negligible effect
on the phonon frequencies. This assumption is supported by recent
Raman measurements in 2-SL MBT samples with inherent anti-site
disorder, since the measured phonon frequencies are in agreement with
density functional calculations for pristine samples~\cite{Choe2021}. 

Next, we introduce disorder in our calculations for the exchange
interactions. The 
anti-site disorder is assumed to be an interchange of Mn with Bi(Sb) 
elements between the Mn layer and Bi(Sb) layers. This is consistent 
with recent experiments \cite{wimmer2021mnrich}. Anti-site disorder effects were found to 
have a quantitatively important effect on the exchange interaction in these materials. Disorder effects are treated using a coherent potential
approximation (CPA) as it is implemented within multiple scattering
theory~\cite{Gyorffy1972}. We show our results in Fig.
\ref{fig:fig4}, where we consider $5 \%$ anti-site disorder, which is
a realistic concentration in most of the known MXT
samples~\cite{PhysRevX.11.021033,otrokov_prediction_2019,wimmer2021mnrich}.
In general, anti-site disorder favors a ferromagnetic interlayer
coupling. The main reason for this is that Mn moments in Bi(Sb) layers
favour a long-range ferromagnetic coupling between the septuple
layers~\cite{Vergniory2014}. Also the reduction of magnetic moments in
Mn layers diminishes the antiferromagnetic coupling.  At zero
displacement, a finite amount of disorder can weaken the effective
exchange interaction, leading to weaker electric fields necessary to
drive the transition. In Fig.  \ref{fig:fig4}, the concentration we
consider leads to a disorder-induced ferromagnetic ground state.


In the previous sections, we established theoretically the possibility
to tune the interlayer magnetic order from antiferromagnetic to
ferromagnetic in 2-SL MXT samples using light in resonance with the
phonons. In this section, we demonstrate that a topological transition
accompanies such a light-induced magnetic transition.

The topology in MBT is rich. In bulk MBT the magnetic structure is
invariant with respect to time-reversal and half a lattice translation
symmetries. This leads to a $\mathbb{Z}_2$ topological classification,
with $\mathbb{Z}_2=1$~\cite{otrokov_prediction_2019}. In the thin-film
limit, the topology depends in the number of
SLs~\cite{otrokov_unique_2019}. For example, 1-SL MBT is predicted to
be a FM trivial insulator, with Chern number $C=0$. 2-SL, 4-SL and
6-SL MBT present a zero plateau QAH, with C=0 in the AFM phase and
$|C|=1$ in the FM phase. Odd-layer (3-, 5- and 7-SL) MBT is predicted
to be in a $|C|=1$ QAH insulating state. Experimentally, the QAH state
has been observed in 5-SL MBT at 1.4 kelvin~\cite{Deng_2020} and a zero
Hall plateau -characteristic of an axion insulating state- in 6-SL
MBT~\cite{Liu2020mbt}.

\begin{figure}
  \centering
  \includegraphics[width=9.2cm]{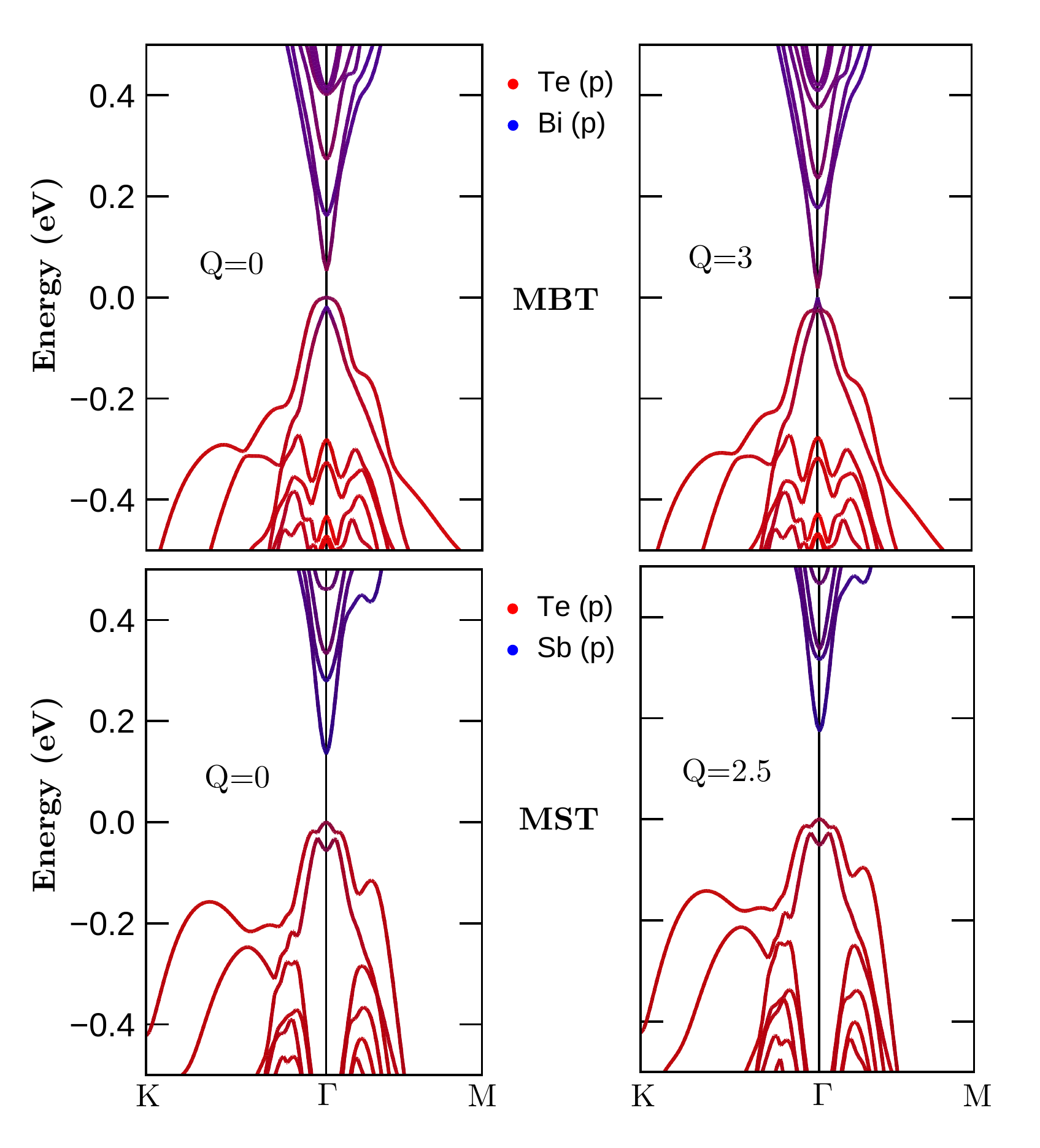}
  \caption{(Color online.)  Band structure with projected $p$ states for
    2SL-MXT in the FM state. In all cases (FM static ($Q=0$) and FM
    out-of-equilibrium ($Q \neq 0$)), we find that the bands are
    inverted.}
\label{fig:fig5}
\end{figure}

We study the topology of 2-SL MBT as a function of the lattice
displacements by examining the electronic band structure and the
projection of the $p$ X=Bi, Sb and Te states. The band inversion
serves as an indicator of the topological nature of the material
within topological band theory~\cite{RevModPhys.88.021004}. Our
results are summarized in Fig. \ref{fig:fig5}. In the equilibrium
configuration (left panels with $Q=0$) with FM order, both 2SL-MBT and
2SL-MST exhibit the expected band
inversion~\cite{otrokov_prediction_2019,wimmer2021mnrich}. For the
out-of-equilibrium distorted structures (right panels), FM order is
preferred as we showed in the previous sections. We find that the band
inversion is present, which indicates the topological nature of the
new laser-induced structures.


This work studied the effect of terahertz light pulses in resonance
with infrared phonons in the magnetic and topological order of 2-SL
MXT samples theoretically. We found that moderate laser intensities,
attainable in current experimental setups, can induce non-linear
dynamics in the Raman breathing mode. The time-average of these
dynamics leads to effective lattice distortions that separate apart
the SLs, effectively increasing the distance between magnetic atom
planes. Using first-principles methods, we found that the new
non-equilibrium lattice configuration can favor ferromagnetic order.
Furthermore, the transition between antiferromagnetic and magnetic
order can be tuned via anti-site disorder. We showed that the magnetic
change is accompanied by a topological transition, as diagnosed by a
band inversion as a function of phonon amplitude. Thus, our
theoretical work demonstrates the possibility of achieving a
sought-after magnetic-topological transition in 2-SL MXT samples
experimentally. Such transition in both 2-SL MBT and MST establishes a
broader materials trend, which could be applied to other van der Waals
magnetic topological materials.

\section*{Acknowledgements} 

We thank Michael Vogl for useful discussions. This research was
primarily supported by the National Science Foundation through the
Center for Dynamics and Control of Materials: an NSF MRSEC under
Cooperative Agreement No. DMR-1720595, with additional support from
NSF DMR-1949701 and NSF DMR-2114825. This work was performed in part
at the Aspen Center for Physics, which is supported by National
Science Foundation grant PHY-1607611. A.L. acknowledges support from
the funding grant: PID2019-105488GB-I00. M. R-V. was supported by LANL
LDRD Program and by the U.S. Department of Energy, Office of Science,
Basic Energy Sciences, Materials Sciences and Engineering Division,
Condensed Matter Theory Program. M.G.V. thanks support from the Spanish Ministry of Science and
Innovation (grant number PID2019-109905GB-C21) and Deutsche Forschungsgemeinschaft (DFG, German Research Foundation) GA 3314/1-1 – FOR
5249 (QUAST).

\section*{Supporting Information Available} 

Additional details on the group theory phonon symmetry analysis, character table for the crystal point group, phonon first-principles calculations, and a discussion on the single-particle excitation spectrum.


\providecommand{\latin}[1]{#1}
\makeatletter
\providecommand{\doi}
  {\begingroup\let\do\@makeother\dospecials
  \catcode`\{=1 \catcode`\}=2 \doi@aux}
\providecommand{\doi@aux}[1]{\endgroup\texttt{#1}}
\makeatother
\providecommand*\mcitethebibliography{\thebibliography}
\csname @ifundefined\endcsname{endmcitethebibliography}
  {\let\endmcitethebibliography\endthebibliography}{}

\end{document}